\documentclass[12pt]{article}
\usepackage{feynmp}
\usepackage[top=1in, bottom=1in, left=1in, right=1in]{geometry}
 \pdfoutput=1
\usepackage[font=footnotesize]{caption}
\usepackage{array}
\usepackage{enumerate}
\usepackage{float}
\usepackage{subfig}
\usepackage{multicol}
\usepackage{multirow}
\usepackage{epsfig}
\usepackage{graphicx}
\usepackage{pict2e}
\usepackage{color}
\usepackage{amsmath,amsfonts,amssymb}
\usepackage{bbm}
\usepackage[utf8]{inputenc}
\usepackage[english]{babel}
\usepackage{xspace}

\newcommand{\be}{\begin{equation}}
\newcommand{\ee}{\end{equation}}
\newcommand{\bea}{\begin{eqnarray}}
\newcommand{\eea}{\end{eqnarray}}

\begin{document}

\title{ Hypercharge Axion and the Diphoton $750$ GeV Resonance}
\date{ }
\author{Ido Ben-Dayan,
Ram Brustein \\ {\small Department of Physics,
Ben-Gurion University, Beer-Sheva 84105, Israel} \\
{\small email: ido.bendayan@gmail.com, ramyb@bgu.ac.il}
}

\maketitle

\begin{abstract}
{The CMS and ATLAS reports on a possible excess of diphoton events at $750$ GeV are the cause of great excitement and hope. We show that a pseudoscalar axion coupled to the topological density of hypercharge, suggested in the past by Brustein and Oaknin as a candidate for inducing baryogenesis, can explain the signal in the diphoton channel. The hypercharge axion (HCA) can also decay to $Z\gamma$ and $ZZ$. The expected number of events in these channels is too small to have been detected, but such decays should be observed in the future.
The HCA can be produced via vector boson fusion (VBF) or via associated production (AP). The latter should be manifested by a characteristic decay to 3 photons which should be observed soon.  We adapt the previous analysis of possible detection of the HCA at the LHC  by Brustein and Oaknin and by Elfgren to the case that the mass of the HCA is 750 GeV and find the expected cross section and decay width in terms of the HCA coupling.  We find that the expected cross-section fits well the observed number of excess events and that there is  tension between the measured cross-section and  the reported partial decay width. The tension is caused because for both VBF and AP the production cross-section and the decay width depend linearly on each other. We therefore expect that if the new resonance is indeed the HCA, either its width should be significantly smaller than the reported width or the production cross-section should be significantly higher than the reported one.}
\end{abstract}

\section{Introduction}

Recent reports by ATLAS and CMS present some hints of an excess of events in the diphoton channel at varying statistical significance \cite{atlas,cms}.
If verified, it indicates the discovery of new physics beyond the Standard Model of Elementary Particles (SM). Such a fascinating possibility has stirred the community \cite{Kobakhidze:2015ldh,Dev:2015isx,Cao:2015pto,Hernandez:2015ywg, Huang:2015evq, Modak:2016ung,Franceschini:2015kwy, DiChiara:2015vdm, Aloni:2015mxa,Chala:2015cev,Ellis:2015oso,Gupta:2015zzs,Csaki:2015vek,Csaki:2016raa, Fichet:2015vvy, Fichet:2016pvq,Pilaftsis:2015ycr, Low:2015qep,Cline:2015msi,Potter:2016psi,Ghorbani:2016jdq}. Specifically, several analyses argue that such a discovery implies several additional new particles, strong dynamics or exotic particles that must come along.
Pressing challenges for explaining the present data are i) Explain the observed excess in the diphoton channel, while such a signal is absent in other channels (for example, $Z\gamma$ or $ZZ$.)  ii) Explain the absence of the signal in RUN I of the LHC  \cite{CMS:2015cwa,Aad:2015mna} and iii) Explain how the width of the observed excess is in accord with the measured cross-section.

The Hypercharge Axion (HCA) was proposed in 1999 by Brustein and Oaknin \cite{BO1} as part of a theory explaining the matter-antimatter asymmetry of the Universe (See also, \cite{dine,Guendelman}.) Subsequently, possible collider signatures were analyzed in \cite{BO2} and various embeddings of the model in a broader framework were discussed in \cite{BO3}. We show that the Hypercharge Axion (HCA) can explain the new resonance by adding only a \textit{single} new pseudoscalar particle to the SM. This is perhaps the minimal setup, because the HCA couples at tree-level only to photons and $Z$-bosons, but not to any other SM particles. The branching ratio of the decay channels of the HCA to photons and $Z$'s is fixed by the Weinberg angle such that the diphoton channel is enhanced compared to the $Z\gamma$ and the $ZZ$ channels. Measuring the branching ratios of the different channels can be used to verify the model as further events are accumulated.

The HCA can be produced at the LHC via vector boson fusion (VBF) or via associated production (AP) with another photon or $Z$. The AP production channel at the LHC was studied by Brustein and Oaknin \cite{BO2}.  Its key signature is  a characteristic event with a 3 photon final state which should be observed soon. The VBF production channel and the possible detection of the HCA at ATLAS was studied by Elfgren \cite{Elfgren:2000ch}.  By adapting the results of the analyses for AP and VBF, we find the production cross-section and the width for a 750 GeV HCA.
Previously, limits on the mass and coupling of the HCA were placed by LEP. The OPAL experiment has put an upper bound on the total cross section of the process $\sigma(e^+ e^- \rightarrow \gamma X) B(X \rightarrow 2\gamma) < 0.04\, pb$, for $30\, GeV < m_X <150\, GeV$, at energies $\sqrt{s}$ up to $189$ GeV \cite{Abbiendi:2002je}.

For both the AP and VBF production channels, the cross-section and the decay width are linearly related, because both quantities are determined by the same coupling. Therefore, a larger decay width implies a larger cross-section and vice-versa.  
 Additional interactions of the resonance exacerbate the puzzle of no signals in alternative channels. The AP and VBF production channels must be present if the resonance can decay to two photons. We thus believe that the tension between the production cross-section and the width has to be faced by any model attempting to describe the new resonance.

The compatibility issue of Run II with RUN I originates from the absence of excess of events, for $\sqrt{s}=8\, TeV$ and $\sigma \simeq 20 fb^{-1}$ \cite{CMS:2015cwa,Aad:2015mna}, compared to the present excess when only $5.8\, fb^{-1}$ have been collected by both ATLAS and CMS at $\sqrt{s}=13\, TeV$. For the two runs to be compatible, the signal cross-section has to increase at least by a factor of a few. However, the compatibility is rather sensitive to the exact distribution of the initial partons, and such a factor could be reasonable \cite{Franceschini:2015kwy}. We will not address this issue further, but wish to note the following. In hadron colliders, we may assign a typical parton-collision-energy as $\tilde s\sim 0.1\sqrt s=800\, GeV$. Then, the production process of an X particle whose mass is $750\, GeV$ is expected to be suppressed due to threshold effects.

In the next section we describe the general setup of the model. In section $3$ we address and analyze the production and decay of a 750 GeV HCA to two photons, and mention some final comments in section $4$.

\section{The HyperCharge Axion}

Pseudoscalar fields appear in several contexts, e.g. \cite{Peccei:2006as,Svrcek:2006yi}. Such fields $X$ are protected by a shift symmetry $X\rightarrow X+c$. Hence, they typically have only perturbative derivative interactions and therefore vanishing potential. Their potential is generated by  non-perturbative effects, many times keeping some residual shift symmetry intact. Consider a pseudoscalar singlet of the SM, $X$. As such, non-perturbative effects will generate a potential of the form $V(X/f)$ where $f$ is the axion decay constant, and periodicity $X\rightarrow X+2\pi f$. Coupling the axion to the Hypercharge topological density, and considering for simplicity  $V(X/f)=\Lambda^4\left(1-\cos(X/f)\right)$, the SM Lagrangian is supplemented by:
\be
 \label{lag}
{\cal L}_{X}=\frac{1}{2}(\partial_{\mu}X)^2- \Lambda^4\left(1-\cos \frac{X}{f}\right)+\frac{1}{8 M}\ X \epsilon^{\mu\nu\rho\sigma}
 Y_{\mu\nu} Y_{\rho\sigma},
 \ee
where $Y_{\mu\nu}$ is the $U(1)_Y$ hypercharge field strength, and the coupling $\frac{1}{M}$ has units of mass${}^{-1}$. Notice that the last term is a dimension $5$ operator, and hence we are discussing an effective theory with some cutoff. The cutoff can be, in general, different than $M$, since unlike the QCD axion, the HCA coupling to the abelian hypercharge topological density does not generate a potential for the HCA. Hence, the HCA must therefore get its mass from some additional sector \cite{BO3}. As a consequence the scales $M$ in the hypercharge sector and in the mass generation sector $\Lambda,f$ (\ref{lag}) are not related. The axion can be displaced from its minimum in the early universe with interesting consequences, but for now we assume that it has reached its global minimum at $X=0$, where its mass is given by $m_X^2\equiv\Lambda^4/f^2$. Despite these general arguments, we consider for simplicity $M\sim f$.
As we will demonstrate, such simplicity makes a large width $\Gamma/m_{X}\sim {\cal O} (1\%)$, technically natural. Moreover, it allows simple restoration of the shift symmetry $X\rightarrow X+c$ at $f\rightarrow \infty$ keeping the symmetry breaking under control.

Since it couples to two neutral vector bosons, the HCA can be produced via two tree level mechanisms: {\bf a)} VBF and specifically, photon fusion: $q+\bar q\rightarrow q+\bar q +\gamma +\gamma \rightarrow q+\bar q+X$ and
{\bf b)} AP, $q+\bar q \rightarrow Z^*/\gamma^*\rightarrow Z/\gamma \,X $ with three gauge bosons as its final state, and $Z^*/\gamma^*$ denote a virtual particle. Both VBF and AP are depicted in Figure 1.

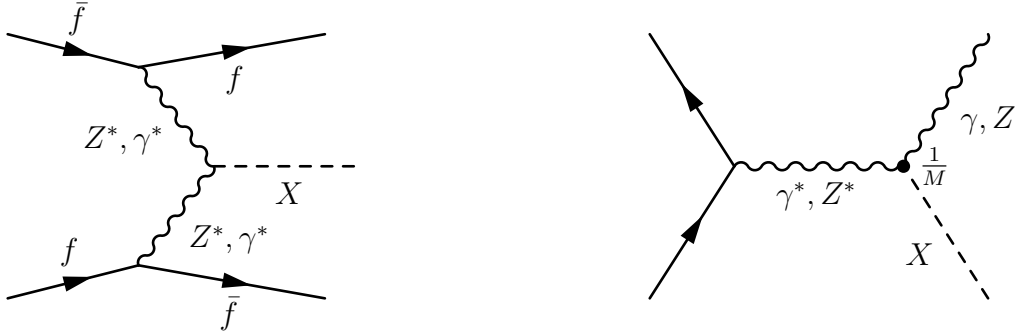
\begin{figure}
[t]
\begin{center}
\begin{fmffile}{vbfF}
\begin{fmfgraph*}(150,100)
\fmfleft{ip,il}
\fmfright{oq1,oq2,d1,oq3,d2,d3,ol}
\fmf{fermion,label={$f$}}{ip,vp}
\fmf{photon,label={$Z,,\gamma$}}{vp,vq}
\fmf{dashes,label=$X$}{vq,oq3}
\fmf{phantom}{ip,vp}
\fmf{fermion,label={$\bar f$}}{vp,oq1}
\fmf{photon,label={$Z,,\gamma$}}{vl,vq}
\fmf{fermion,label={$\bar f$}}{il,vl}
\fmf{fermion,label={$f$}}{vl,ol}
\fmf{phantom}{il,vl}
\end{fmfgraph*}
\end{fmffile}
\hspace{1in}
\begin{fmffile}{vbfI}
\begin{fmfgraph*}(150,100)
\fmfleft{i1,i2}
\fmfright{o1,o2}
\fmf{fermion}{i1,v1,i2}
\fmfdot{v2}
\fmflabel{$\frac{1}{M}$}{v2}
\fmf{dashes,label=$X$}{o1,v2}
\fmf{photon,label=$\gamma,,Z$}{v2,o2}
\fmf{photon,label=$\gamma^*,,Z^*$}{v1,v2}
\end{fmfgraph*}
\end{fmffile}
\end{center}
\caption{Production of the HCA via vector boson fusion (left) and associated production (right). $Z^*,\gamma^*$ are virtual bosons.}
\end{figure}

\begin{figure}
[H]
\begin{center}
\begin{fmffile}{vbfJ}  
\begin{fmfgraph*}(240,240)
\fmfright{ip,il}
\fmfleft{oq1,oq2,d1,oq3,d2,d3,ol}
\fmf{photon,label={$Z,,\gamma$}}{vp,vq}
\fmf{dashes,label=$X \hspace{1in}\frac{1}{M}$}{vq,oq3}
\fmfdot{vq}
\fmf{phantom}{ip,vp}
\fmf{photon,label={$Z,,\gamma$}}{vl,vq}
\fmf{phantom}{il,vl}
\end{fmfgraph*}
\end{fmffile}
\end{center}
\caption{Decay diagram of the HCA }
\end{figure}
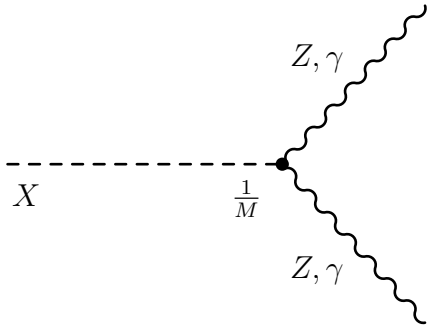
The HCA decays to $\gamma \gamma, \gamma Z, ZZ$, as depicted in Figure $2$. Its decay rate is given by:
\bea
\Gamma(X \rightarrow 2 \gamma) &=& \frac{1}{64\pi} \cos^4 \theta_W
\frac{m_X^3}{M^2}, \\
 \Gamma(X \rightarrow Z \gamma) &=& \frac{2}{64\pi}
 \cos^2\theta_W \sin^2 \theta_W
 \left(\frac{(m_X^2-m_Z^2)^3}{m_X^3} \frac{1}{M^2}\right),
\\
 \Gamma(X \rightarrow 2 Z) &=& \frac{1}{64\pi}
  \sin^4 \theta_W \frac{1}{M^2}(m_X^2 - 4 m_Z^2)^{3/2}.
\eea
Here $m_X$ is the mass of the axion.
For a mass $m_X=750\, GeV$, the branching ratios to the different channels are weakly dependent on $m_X$ and do not depend on the coupling $M$.
The branching ratios are given by
\bea
\label{br}
BR(X \to 2 \gamma) &=& 0.61, \\
BR(X \to \gamma Z) &=& 0.34, \\
BR(X \to 2 Z) &=& 0.05.
\eea

The fact that even in the diphoton channel only order of $\sim 25$ events have been observed by ATLAS and CMS combined, explains the observation in the diphoton channel and the absence in other channels. The $Z\gamma$ channel is suppressed by a factor of $1.8$ compared to $\gamma \gamma$, so one could have hoped for $\sim 15$ events. However most subsequent decays are expected to be swamped by the background. The cleanest detection channel is through the decay  $Z\rightarrow l \bar l$. However, this comes with a price tag, a branching ratio of $6\%$. Thus, we understand why such events were not observed \cite{Elfgren:2000ch}. The $ZZ$ channel is responsible for only about $5\%$ of the decays, so even before discussing the possible backgrounds one would expect at most $1-2$ events.

For both production mechanisms, we expect a two body decay which is not too different from spherical, assuming that for most events, the HCA is produced as a non-relativistic particle.  Of course, being a pseudoscalar, the helicities of the photons have to be correlated appropriately.
In the case of AP we discuss the angular distribution of the associated photon in the last section.

Precision test of the SM are almost blind to the HCA because it is a singlet and couples only to hypercharge in a gauge invariant way.
For example, the additional HCA term at tree level does not affect the oblique parameters $S$, $T$, $U$:
\bea
\alpha T&=&\frac{\Pi_{WW}(0)}{m_W^2}-\frac{\Pi_{ZZ}(0)}{m_Z^2},\cr
\alpha S&=&4c^2s^2\left[\Pi'_{ZZ}(0)+\frac{c^2-s^2}{s c}\Pi'_{Z\gamma}(0)-\Pi'_{\gamma \gamma}(0)\right],\cr
\alpha U&=&4 s^2\left[\Pi'_{WW}(0)-c^2\Pi'_{ZZ}(0)-2s c \Pi'_{Z\gamma}(0)-s^2\Pi'_{\gamma \gamma}(0)\right].
\eea
with $c=\cos \theta_W,\, s=\sin \theta_W$ and $\Pi_{AB},\, \Pi'_{AB}$ denote the self-energy contributions of gauge boson $A,B$ as the external legs of the Feynman diagram.  The contribution to the $T$ parameter vanishes because the HCA coupling terms is proportional to momentum square due to the derivatives in the interaction terms, and the $T$ parameter is evaluated at $p=0$.
The contributions to $S,U$ parameters vanish because the HCA is an $SU(2)$ singlet, so $\Pi_{WW}=\Pi'_{WW}=0$ and the other $\Pi'_{AB}$ cancel out because of the $U(1)$ invariance.

\section{Production Cross-section and Decay Width to Photons of a 750 GeV HCA}

As mentioned, the production cross-section via AP and VBF were studied previously. Despite the fact that the studies were conducted long time ago, the parameters of the LHC were known at that time and the estimates are expected to be rather accurate. We hope to revisit and update them in a future publication.

\subsection{Theoretical Considerations}

\subsubsection{Cross-section}

The cross-section for AP was calculated in \cite{BO2} and evaluated numerically for a 14 TeV LHC. Because the cross-section scales as $1/M^2$, the results could be read off Figure 7 of \cite{BO2}, for $m_X=750 GeV$,
\be
\sigma{(pp\to X \gamma/Z)}=12\, fb \left(\frac{TeV}{M}\right)^2.
\ee
The cross-section for VBF (essentially, photoproduction)  was calculated in \cite{Elfgren:2000ch} and evaluated numerically for a 14 TeV LHC. Again, the results can read off Figure 3.2 of \cite{Elfgren:2000ch},
\be
\sigma{(pp\to X )}=2.6\, fb \left(\frac{TeV}{M}\right)^2.
\ee

The total production cross-section times the branching ratio to two photons from Eq.~(\ref{br}) is then given by
\be
\label{txs}
\sigma{(pp\to X )} BR (X\to 2\gamma) \simeq 9\, fb \left(\frac{TeV}{M}\right)^2
\ee
This result should be viewed as a reasonably accurate estimate to the actual value. However, to get a more accurate result that can be compared with a future measurement of the cross-section one needs to redo the numerical evaluation of the cross-section for a 13 TeV LHC and with updated PDF's. We expect that results, taking into account more precise PDF's, the reduction of energy  from 14 TeV to 13 TeV, etc.,  will differ by an order of 10\% at most from the previous results.

\subsubsection{Decay width}
For a 750 GeV  HCA:
\be
\label{ratio}
\Gamma(X\rightarrow 2\gamma)=1.25\, GeV \left(\frac{m_X}{750\, GeV}\right)\left(\frac{TeV}{M}\right)^2.
\ee
Hence, a large width in the GeV range is natural in the case of $m_X \lesssim M \sim TeV$. If there is a larger hierarchy between the scales, the width can be extremely narrow. A greater challenge is the possibility that the ATLAS width, $\Gamma=45\, GeV$ survives additional analysis and data collection. A width $\Gamma=45\, GeV$ requires the partial width above to be  $\Gamma(X\rightarrow 2\gamma)=27\, GeV$, i.e. $\left(\frac{TeV}{M}\right)^2\sim 22$.

Theoretically, a large width can be accommodated. Let us consider the axion potential $V(X/f)$. As explained, the potential is generated by non-perturbative effects, and therefore, in principle, is not associated with the mass scale $M$. So $m_X^2=\partial_X^2V$ can be larger than $M^2$. Let us adopt the cosine potential form \eqref{lag}, and assume that $M=f$.
In such a case we have, $m_X^2=\Lambda^4/f^2$. Substituting into \eqref{ratio} gives:
\be
\Gamma (X\rightarrow 2\gamma)\simeq 2.25\, GeV \left(\frac{m_X}{750\, GeV}\right) \left(\frac{\Lambda}{f}\right)^4.
\ee

Due to the quartic dependence, to obtain a large width, all that is needed is $\Lambda \gtrsim f$, so the large width is natural in the HCA model, and no small or large numbers need to be introduced.  More importantly, the analysis here is preformed by using the simplest possible axion model. Much larger hierarchies than $\mathcal{O}(1)$ numbers are easily obtained once two or more different axions are considered, \cite{Ben-Dayan:2014zsa,Ben-Dayan:2014lca,Tye:2014tja}.

Another possibility to achieve a width enhancement is by considering $N$ copies of the axion. In that case the single axion width is multiplied by $N^2$ giving
\be
\Gamma(X\rightarrow 2\gamma)=N^2 \times 2.25\, GeV \left(\frac{m_X}{750\, GeV}\right)\left(\frac{m_X}{M}\right)^2.
\ee
So to get an enhancement factor  $\sim 12$, three-four copies, $N=3-4$ and $M \sim 750\, GeV$ are sufficient.

While the large width can be accommodated, loop corrections could present a problem, especially since the theory we consider is an effective theory. So, one should worry whether contribution from loops will be as large as the tree level ones, invalidating our analysis. This concern is certainly relevant in the case we are discussing where $m_X\sim M$. However, two effects may relax the concern. First, the loop contributions are suppressed by factors of $1/16\pi^2$. So while the 1-loop might contribute, the higher order loops will still be negligible even in the $m_X\sim M$ case. Second, as we have explained, the axion potential is generated from non-perturbative effects, and as such, naive dimensional analysis does not necessarily hold.
To summarize, the broad width reported by ATLAS, that seems hard to accommodate in many models can be generated in the HCA scenario.

However, as we now show, considering the larger width $\Gamma \sim 45\, GeV$ and the number of events measured $\sim 25$, induces a tension between the value of the measured cross-section and the theoretically calculated one. 
Given that the number of events is more reliable than the fit to the width, and CMS pull towards a narrower width, we expect future data to narrow the width down, which will give also a ``healthier" model from the theoretical point of view given the above considerations.

\subsection{Comparison with Experiment}

The two experiments do not report yet numerical values $\sigma_X\, BR(X\to 2\gamma)$ for the excess. However, the values can be read from the corresponding graphs. As a conservative estimate,  we adopt a value $\sigma_X\, BR(X\to 2\gamma) \sim 7 \pm 5 fb$. The total number of observed excess events is about 25 with a large uncertainty and the total integrated luminosity of both CMS and ATLAS is $5.8 fb^{-1}$, taking into account some losses due to imperfect acceptance and the branching ratio, the two estimates seem to be in rough agreement. These estimates are also in agreement with other estimates \cite{Franceschini:2015kwy,DiChiara:2015vdm,Aloni:2015mxa,Ellis:2015oso,Gupta:2015zzs,Csaki:2015vek,Csaki:2016raa},
 which give estimates of $\sigma_X\, BR(X\to 2\gamma) \lesssim 10 \,fb$.

Our estimate of $\sigma_X\, BR(X\to 2\gamma) \sim 9 fb \left(TeV/M\right)^2$ in Eq.~(\ref{txs}) fits very well the estimates of the measured quantity, provided that $M \sim TeV$. In this case, we see that the AP and VBF processes saturate the expected number of events and therefore suffice as the sole production mechanisms.

As for the width of the resonance $\Gamma_X$, ATLAS has reported a mild preference for a broad width $\sim 45\, GeV$ while CMS mildly prefers a narrower width of about $20\, GeV$. This is the smallest width that CMS considered in the search.  From Eq.~(\ref{ratio}) we see that the ATLAS result requires $M \sim 215 GeV$ while the CMS result requires $M \sim 320 GeV$. So there is some tension with the above requirement of $M\sim TeV$. It is likely that the width estimate is more uncertain than the estimate of the number of observed events and therefore the estimated cross-section. We therefore expect that the measured width will eventually be smaller, of the order of a few GeV.

\section{Concluding Remarks}

In brief, we proposed the HCA  with $m_X=750 \, GeV$ as an explanation for the recent observed diphoton excess at the LHC \cite{atlas,cms}. The cross section for production of the excess diphoton events matches the measured one for an HCA coupling scale $M\sim TeV$ and we find that a large decay width is quite natural in the HCA model.

The actual reach of the LHC for detection of the HCA will be determined by the backgrounds. The SM background to the three photons signature is mainly a pure QED process whose differential cross-section is strongly peaked along the forward and backward directions. The three photons events coming from near resonance AP and decay of the HCA particle have a more isotropic distribution. The angular dependence of the differential cross sections for AP is $(1 + cos^2 \theta_\gamma)$ where $\theta_\gamma$ is a polar angle with the beam axis.  This is the most immediate test of the HCA model. For VBF, one has to observe the forward jets of the production partons. Moreover, since the same vertex is in charge of both VBF and AP, it seems that any alternative explanations to the diphoton resonance should also predict a signal of three gauge bosons via AP, that should be corroborated by the experiment soon.

To confirm that the $X$ particle actually couples to hypercharge, and not merely to electromagnetic topological number density,  the other two neutral gauge bosons signatures have to be detected in the appropriate ratios. Since the mass of the HCA is known, additional cuts on the invariant mass of each pair of produced bosons could be imposed. This would drastically improve the signal to background ratio in these other channels. According to \cite{Elfgren:2000ch}, without using the decay modes to jets of the Z-bosons, as many as three years of high-luminosity LHC operation might be needed for discovery.

The model we presented is rather minimal and the observed resonance explained by an effective field theory in which the only tree level coupling is between the HCA and the neutral gauge bosons. Considering additional interactions, such as direct coupling to gluons or to other particles we expect that either the branching ratio to photons should be considerably smaller than for the HCA, and/or the cross section should be considerably larger.

{\bf \large Acknowledgements} \\
We thank Liron Barak, Erez Etzion, Gian Giudice, Yuval Grossman, Yevgeny Katz, David Oaknin and Gilad Perez for useful discussions.

\end{document}